\documentclass[conference]{IEEEtran}
\IEEEoverridecommandlockouts
\usepackage{cite}
\usepackage{amsmath,amssymb,amsfonts}
\usepackage{algorithm}
\usepackage{algpseudocode}
\usepackage{graphicx}
\usepackage{textcomp}
\usepackage{xcolor}
\usepackage[normalem]{ulem}
\usepackage{multicol}
\usepackage{lipsum}
\usepackage{multirow}
\usepackage{xspace}
\usepackage{caption}
\usepackage{subcaption}
\usepackage{amsmath}
\usepackage{amsthm}
\newtheorem{theorem}{Theorem}
\usepackage{paralist}
\usepackage[inline]{enumitem}
\usepackage[bottom]{footmisc}
\setlist[enumerate,1]{label=\textit{\alph*)}}
\usepackage{environ,censor,calc}
\NewEnviron{mysout}
 {\soutrefunexp{\BODY}}
\newlength\nextcharwidth
\makeatletter
\renewcommand\@cenword[1]{%
  \setlength{\nextcharwidth}{\widthof{#1}}%
  \censorrule{\nextcharwidth}%
  \kern -\nextcharwidth%
  #1}
\makeatother
\newcommand\soutref[1]{\censorruledepth=.55ex\xblackout{#1}}
\newcommand\soutrefunexp[1]{\expandafter\soutref\expandafter{#1}}

\usepackage{hyperref}
\hypersetup{pdfstartview=FitH,
  pdfpagelayout=SinglePage,
  bookmarksnumbered=true,
  bookmarksopen=true,
  colorlinks=true,
  linkcolor=teal,
  urlcolor=teal,
  citecolor=teal}

\newcommand{\namex}{RICS\xspace}

\usepackage{algpseudocode}
\algtext*{EndWhile}
\algtext*{EndIf}
\usepackage{varwidth}

\def\BibTeX{{\rm B\kern-.05em{\sc i\kern-.025em b}\kern-.08em
    T\kern-.1667em\lower.7ex\hbox{E}\kern-.125emX}}
\begin{document}

\title{Routing for Intermittently-Powered Sensing Systems}

\author{\IEEEauthorblockN{Gaosheng  Liu}
\IEEEauthorblockA{\textit{Vrije Universiteit Amsterdam} \\
Amsterdam, Netherlands \\
g.s.liu@vu.nl
}
\and
\IEEEauthorblockN{Lin Wang}
\IEEEauthorblockA{\textit{Vrije Universiteit Amsterdam} \\
Amsterdam, Netherlands \\
lin.wang@vu.nl
}
}


\maketitle

\begin{abstract}

Recently, intermittent computing (IC) has received tremendous attention due to its high potential in perpetual sensing for Internet-of-Things (IoT). 
By harvesting ambient energy, battery-free devices can perform sensing intermittently without maintenance, thus significantly improving IoT sustainability. 
To build a practical intermittently-powered sensing system, efficient routing across battery-free devices for data delivery is essential. 
However, the intermittency of these devices brings new challenges, rendering existing routing protocols inapplicable.

In this paper, we propose \namex, a new routing scheme tailored for intermittently-powered sensing systems. 
\namex features two major designs to combat the intermittency challenge, with the goal of achieving low-latency data delivery on a network built with battery-free devices. 
First, \namex incorporates a fast topology construction protocol for each IC node to establish a path towards the sink node with the least hop count. 
Second, \namex employs a low-latency message forwarding protocol, which incorporates an efficient synchronization mechanism and a novel technique called pendulum-sync to avoid time-consuming repeated node synchronization. 
Our evaluation based on an implementation in OMNeT++ and comprehensive experiments with varying system settings shows that \namex can achieve orders of magnitude latency reduction in data delivery compared with the state-of-the-art.  

\end{abstract}


\section{Introduction}

Intermittent computing (IC) has received increasing traction recently considering its high potential in sustainable Internet-of-Things (IoT) applications~\cite{2017-snapl-survey,2020-asplos-chrt,2020-asplos-tics,2020-imwut-bfgame,2020-imwut-bfree,2022-mobisys-bluetooth,2022-pldi-wario,2021-csur-sync,2020-sensys-bfsensing,2021-weee-cic,2022-tecs-camaroptera,2022-nc-underwater,2019-asplos-genesis}.
IC relies on \emph{battery-free} low-power devices, which harvest ambient energy from the environment (e.g., light, radio-frequency signals, and kinetic energy) and accumulate the energy in capacitors to function~\cite{2020-imwut-bfgame,2017-imwut-bfphone}. 
These devices require zero maintenance once deployed and are environment-friendly attributed to the absence of batteries which typically contain hazardous chemicals. 
Recently, rapid hardware advances~\cite{2018-asplos-energystore,2017-snapl-survey}, and software solutions for enabling continuous task execution under intermittency~\cite{2019-pldi-samoyed,2020-tecs-checkpoint,2022-pldi-wario,2016-oopsla-chain,2019-asplos-genesis,2017-oopsla-alpaca} have laid a solid foundation for IC. 

However, the intermittent nature of IC devices brings challenges to building sensing systems. 
Typically, a sensing system consists of dispersed nodes, which collect data from the environment and send, hop-by-hop, the data to a common, more powerful sink node for complex processing. 
Since not all nodes can communicate with the sink directly (due to limited radio coverage), data has to be \emph{routed} across the nodes. 
The following two specific challenges arise: 
(1) IC nodes need to form a network and each IC node needs to acquire routing information for reaching the sink. 
Such topological information cannot be set in advance since the node distribution can be quite random during the system deployment.
In traditional sensor networks, battery-powered nodes calculate their locations with the aid of auxiliary components like GPS units, which are too power-consuming for IC devices.
(2) Based on the acquired routing information, an IC node has to be synchronized, with respect to the working time (during which the IC device is active), with its next-hop IC node specified in the routing information. 
This is critical for ensuring successful data transfer since the working time of different IC nodes is likely different. 
Despite being a complex task per se, such synchronization has to happen on a per-message, per-hop basis, leading to an extremely time-consuming process that can greatly degrade the routing performance, leading to large data delivery delays.
\emph{Therefore, efficient routing is key to building a practical sensing system with IC devices.}

Existing efforts fall short of addressing the routing problem in intermittently-powered sensing systems. 
On the IC research front, there have been works on exploring the communication issue between a pair of IC devices~\cite{2021-mass-comm,2021-nsdi-find,2022-nsdi-bonito}. 
However, the main focus, so far, is on working time synchronization, which is a major challenge in IC environments~\cite{2021-csur-sync}. 
For example, Find builds a delay model based on a geometric distribution for improving the encountering probability of the working time of two IC devices to facilitate synchronization~\cite{2021-nsdi-find}. 
Recently, a Bluetooth Low Energy (BLE) implementation has been presented for IC devices without protocol specification changes~\cite{2022-mobisys-bluetooth}.
While we will build on these steps, these works do not address the problem of multi-hop routing across a set of IC devices.
Meanwhile, existing routing protocols for traditional wireless sensor networks and delay-tolerant networks, as well as for energy-harvesting wireless sensor networks that harvest energy to recharge sensor batteries, are ill-suited for IC devices since they typically assume more capable nodes (e.g., powered by batteries), often also equipped with geo-location devices, which allow for complex routing protocols to be implemented~\cite{2012-csur-wsn,2018-tsn-ehwsn}. 
Additionally, due to the absence of intermittency, these protocols do not account for the working time synchronization overhead, which is critical to the performance of IC sensing systems. 
\emph{To the best of our knowledge, none of the existing works fully address the routing problem in intermittently-powered sensing systems.}

In this paper, we propose a new routing scheme called \namex, tailored for \underline{R}outing in \underline{IC} \underline{S}ensing systems. 
The goal of \namex is to ensure efficient data transfer across IC nodes so that the data (e.g., sensor readings) generated by each IC node can be delivered to the sink node as timely as possible. 
\namex achieves this goal by featuring the following two specific designs.
First, we propose an \emph{efficient topology construction} protocol where each IC node decides its best next hop for passing messages toward the sink node.
Our proposed protocol starts from the sink node and disseminates and updates the minimum hop count of each node to the sink iteratively, accounting for the intermittency nature of the IC devices. 
Our protocol ensures finding the path for each node to reach the sink node with the least hop count, minimizing the data delivery delay. 

Second, we propose a \emph{low-latency message forwarding} protocol based on a technique called pendulum-sync.
Given a path from an IC node to the sink produced in topology construction, each of the nodes on the path has to synchronize its working time with its next hop before its communication can start. 
This leads to significant time overhead in data delivery if such synchronization is required for each individual message along the path. 
We observe that the offset between the working time of neighboring IC nodes on the path can stay static for a relatively long period of time and, thus, the IC node can cache the offset to avoid the expensive repeated synchronization.
We propose pendulum-sync, a technique that allows the IC node to swing forth its working time (based on the cached offset) to match with its next hop for communication and swing back when the communication is completed. 
With such a protocol, \namex eliminates the need for frequent working time synchronization among the IC nodes, thus reducing the delay in message forwarding and data delivery significantly.

Overall, we make the following contributions to this paper.
\begin{enumerate}[label=(\arabic*)]
    \item We propose an efficient topology construction protocol (Section~\ref{sec:topology}) for building the least-hop path for each IC node to the sink node given a set of IC nodes.
    \item We propose a low-latency message forwarding protocol (Section~\ref{sec:routing}) based on a new technique called pendulum-sync, which caches the working time offset between IC nodes and dynamically adapts the working time of each IC node to communicate with its next hop.
    \item We implement \namex in OMNeT++ and perform extensive experiments to validate its performance under various scenarios and settings (Section~\ref{sec:evaluation}). Our results show that \namex achieves orders of magnitude reduction in the data delivery latency compared with the baselines.
\end{enumerate}
Section~\ref{sec:discussion} discusses the limitations of \namex. 
Section~\ref{sec:conclusions} draws the conclusions.
\section{Overall Design}
\label{sec:design}



\subsection{Target Scenario}
We consider an intermittently-powered sensing system built with identical IC devices (denoted as IC nodes from now on).
IC nodes work in a time-slotted fashion where a time \emph{slot} is defined as the length of the working period of the IC node (during which the IC node is powered on, typically short, at ms scale, with energy buffered in a small capacitor). 
The charging time is denoted by $t$ slots considering that the charging period is typically significantly longer than the working period~\cite{2021-nsdi-find,2022-nsdi-bonito}. 
We assume that the working period is long enough for an IC node to perform at least one sending and one receiving operation (to achieve reliability through acknowledgments), which is practical~\cite{2021-nsdi-find}. 
The time slots of different IC nodes are aligned with existing mechanisms like Flync~\cite{2021-nsdi-find}.

We focus on the case where the IC nodes are deployed in a relatively stable environment (e.g., non-shadowed walls of an office), where the charging cycles of these IC nodes are uniform and stay constant in a given period of time, which is the same case studied in recent works (e.g.,~\cite{2019-sensys-shepherd,2021-nsdi-find}). 
For more dynamic scenarios with varying charging cycles of IC nodes, we can simulate this case by delaying the working period of the IC node, allowing all IC nodes to follow the maximum charging time of all IC nodes (estimated conservatively offline), at the cost of degraded efficiency.
The working time of these IC nodes can be arbitrarily distributed in the charging cycle and may be interleaved by a random offset (in $[0,t]$ time slots) from each other due to physical initialization variations~\cite{2021-nsdi-find}. 


\subsection{System Overview}


\begin{figure}[!t]
    \centering
    \includegraphics[width=0.38\textwidth,page=1]{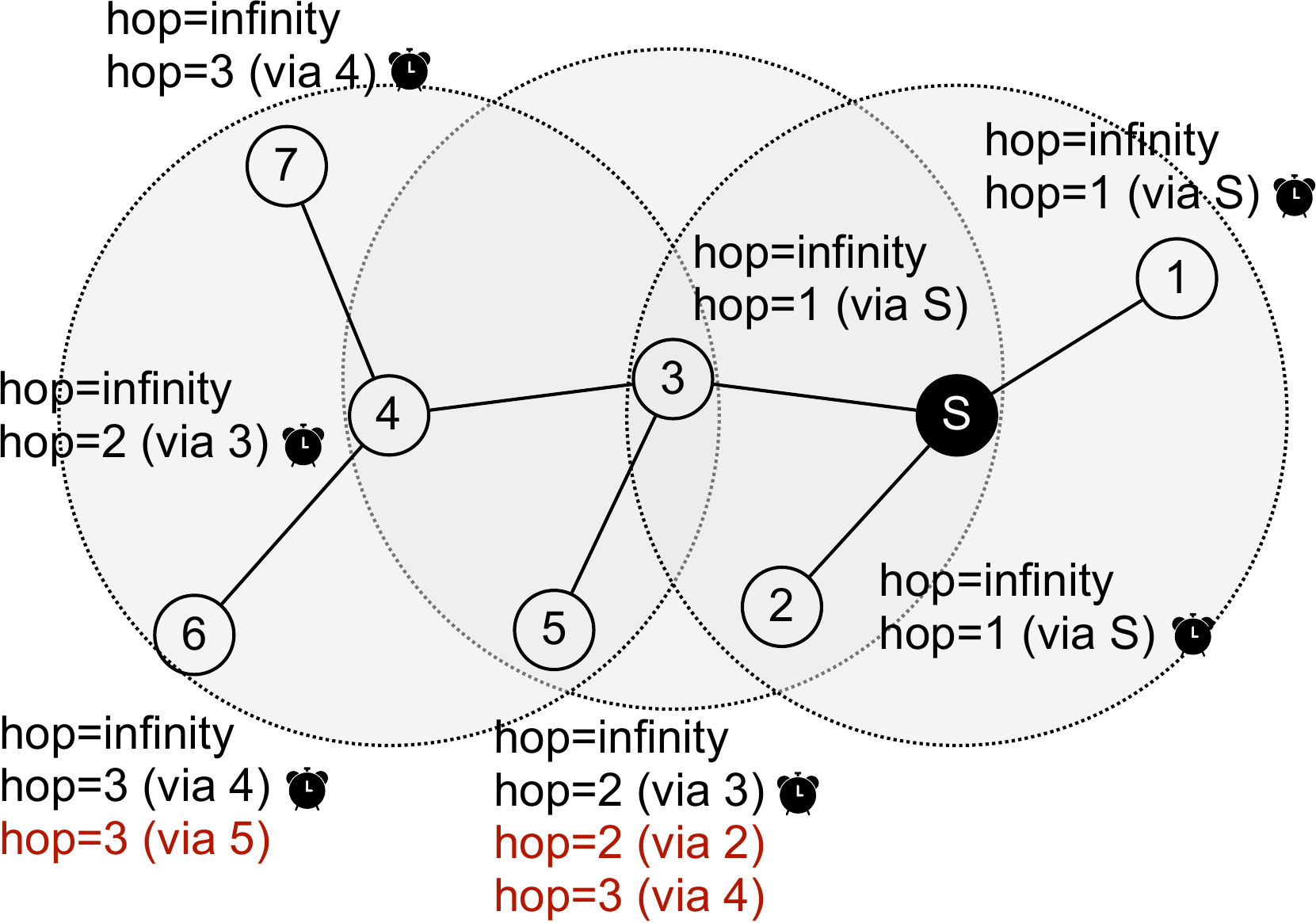}
    \caption{An overview of the \namex topology construction procedure. (\textsf{S} denotes the sink node. Text in red represents the messages ignored by the IC node. The clock icon indicates that the IC node waits before it starts broadcasting.).}
    \label{fig:overview:topo}
\end{figure}

We design \namex---a new routing scheme tailored for intermittently-powered sensing systems. \namex consists of two major components where each IC node starts with the distributed \emph{topology construction} procedure to discover routing information among the nodes in the system and then proceeds with \emph{message forwarding} to send the periodically collected sensor data (at each IC node) to the sink hop-by-hop following the constructed network topology. \namex works on the network layer of the BLE mesh protocol stack~\cite{2018-specification-BLEMesh}.

\subsubsection{Topology construction}
At a high level, the topology construction procedure works as follows:
Essentially, all the nodes in the sensing system adopt a sequential broadcast-wait protocol to discover each other and establish the shortest paths (i.e., paths with the least number of hops) to the sink. 
To that end, each IC node starts with the listening state. 
Upon receiving a hop-count message, an IC node checks if a lower hop count has been found via the source of the received message. 
If so, the IC node updates its next hop to be the message source and broadcasts this new hop count to its neighbors; otherwise, the message is dropped and the IC node remains in the listening state. 
The above procedure is initially triggered by the sink which broadcasts the first hop-count message (with \(hop=0\)) to all the IC nodes within its signal coverage.
Once the above procedure terminates, the network topology converges into a spanning tree with one unique path for each node to reach the sink. 
Using these paths, each node can forward packets to the sink with low latency. 
Figure~\ref{fig:overview:topo} depicts an example of the topology construction procedure with hop count information propagated from the sink outwards. 
We will elaborate on the routing protocol design in Section~\ref{sec:topology}.

\subsubsection{Message forwarding}
Once the topology has been established, the sensing system turns into the message-forwarding phase where data is collected at IC nodes and sent to the sink using the established paths. 
Each IC node generates data (e.g., temperature or humidity readings) and once enough data is produced, a message is created with the monitored data, which needs to be sent to the sink. 
Each IC node maintains its next hop obtained during topology construction. 
However, the working time of an IC node and its next hop is unlikely aligned, so the first challenge to address is the synchronization of working time across every hop. 
To this end, we propose a deterministic working time synchronization mechanism based on observed maximum charging times for an IC node to discover and communicate with its next hop. 
While our proposed mechanism is efficient in normal cases, it can still take $t \times (t-1)$ slots per hop in the worst case, which becomes prohibitively large over multiple hops.

\begin{figure}[!t]
    \centering
    \includegraphics[width=0.38\textwidth,page=2]{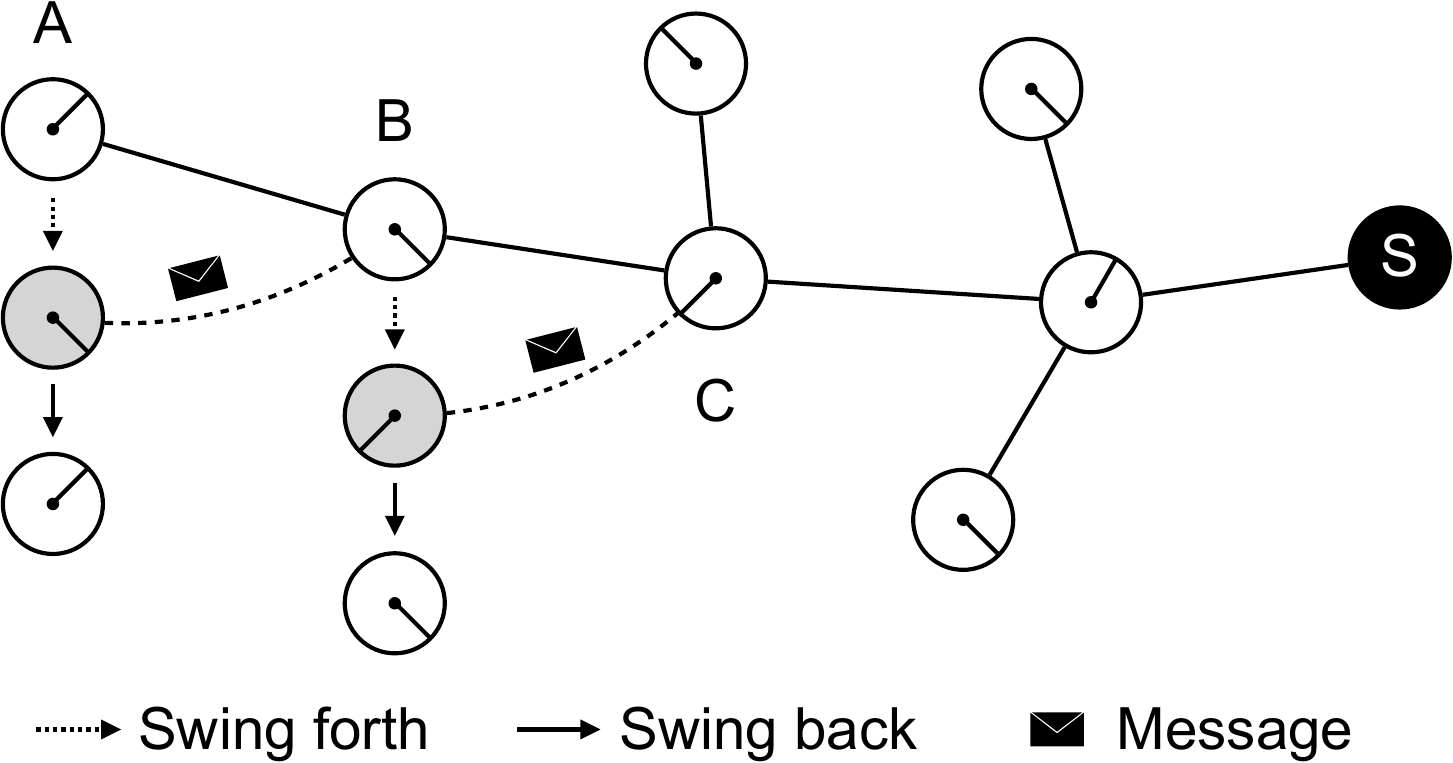}
    \caption{Efficient message forwarding based on the proposed pendulum-sync technique. Each IC node swings forth to align its working period with the next hop to send messages and swings back to receive messages.}
    \label{fig:overview:sync}
\end{figure}

To make message forwarding more efficient, we propose an efficient technique called \emph{pendulum-sync}. 
For each IC node, once the working time synchronization has been done, the sender first swings forth its working time to align with the next hop to send messages and then swings back its working time to its original position to receive messages from upstream IC nodes. 
A simple example is shown in Figure~\ref{fig:overview:sync}. 
In this example, node \textsf{A} first swings forth to match with node \textsf{B} to send the message and then swings back to its initial working time to listen for upcoming messages. \textsf{B} then swings forth to align with \textsf{C} to pass the message down on the path and swings back to its initial working time afterward so \textsf{A} knows how to communicate with it later. 
We will elaborate on the routing protocol design in Section~\ref{sec:routing}.

\section{Routing Topology Construction}
\label{sec:topology}

In this section, we describe our proposed topology construction protocol in detail and discuss its properties. 
The goal of topology construction is to establish routing information (more precisely the next hop for each IC node to take) so that each IC node can reach the sink node with low latency, i.e., the least number of hops in our case. 

Our topology construction design follows a broadcast-wait mechanism where IC nodes propagate the hop count information (i.e., the distance to the sink node) throughout the network. 
The protocol works as follows: 
We start from the sink node which cycles through the charging cycle of the IC node to broadcast its hop count (zero in this case) to all its neighbor nodes. 
Since IC nodes need at most $t$ slots to charge after initialization, the sink will first wait for $t$ slots before broadcasting.
It takes at most $t+1$ slots for the sink node to ensure that all its neighbors have received its broadcast (by cycling through the whole charging cycle of IC devices). 
Therefore, the sink broadcast phase takes $2t+1$ slots in total. 

\begin{figure}[!t]
    \centering
    \includegraphics[width=0.38\textwidth,page=3]{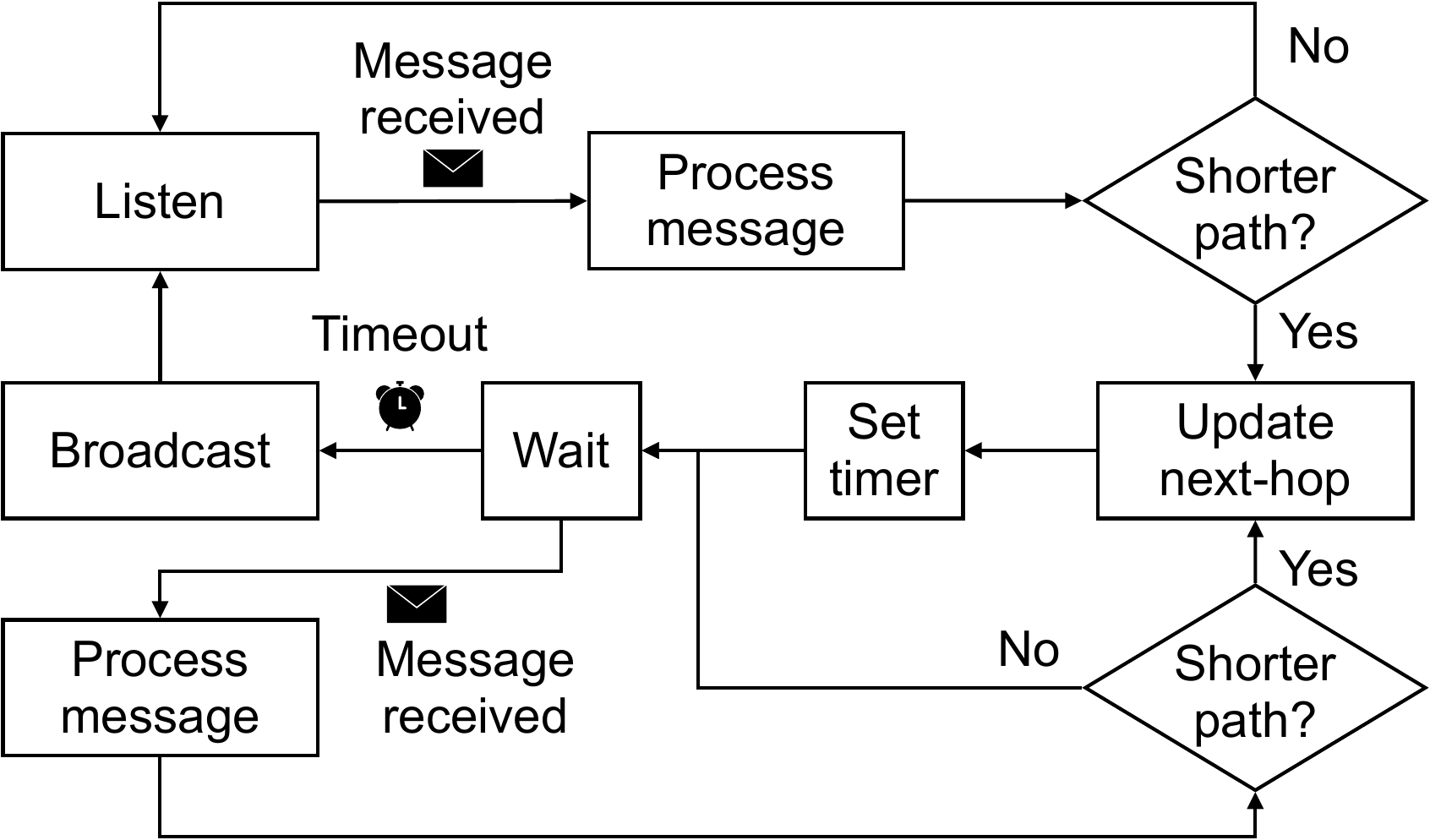}
    \caption{The workflow of an IC node during topology construction, following the flooding-based protocol.}
    \label{fig:topo-workflow}
\end{figure}

The general workflow for each IC node is depicted in Figure~\ref{fig:topo-workflow}.
All the IC nodes in the system execute the following protocol:  
The hop count for all IC nodes is initialized to infinity at the beginning.
Before receiving a broadcast message, an IC node remains in the listening state. 
Upon receiving a broadcast message from the sink (with $hop=0$), the IC node receives the hop values and remaining broadcasting rounds from its sender. Now the receiver updates its saved hop count and continues to wait until its sender finishes its broadcasting then starts to broadcast this new hop count to its neighbors by cycling through its charging cycle, i.e., delaying its working time by one slot in every round until its working time offset returns to its original working time offset. 
This repeats for $t+1$ rounds, each taking $t+1$ slots. 
Further, in each broadcast message, the sender includes its current hop count and the round number $r \in [0,t]$ for this message. 
This round number indicates the broadcast progress of the sender for this particular message. 
Upon receiving a broadcast message from the sender, the receiver will first check if the received hop count plus one is smaller than its current hop count. 
If so, the IC node will apply an update to its hop count with the received hop count plus one and its next hop with the source of the message; otherwise, the received message will be ignored.
In the case of an update, the IC node will check the round information contained in the broadcast message, set a timer for $t-r$ slots, and wait before it starts to broadcast to other nodes. 
This is to ensure that the sender of the broadcast message has already finished broadcasting its message and that all its neighbors (i.e., IC nodes within its coverage) have already received its broadcast message to update their hop counts. 
During the waiting period, if the IC node receives another broadcast message containing an even smaller hop count, the IC node will apply the update and start over the waiting process with the new round information from this new broadcast message. 
The broadcast-wait protocol terminates when all the IC nodes are done with broadcasting.
To deal with collisions, the receiver sends an acknowledgment with a random delay to notify senders, and the sender broadcasts messages with a random delay. If a node cannot receive any hop count value within the maximum waiting time $(t + 1) \times (t + 2) \times N_{maxhop}$, it will switch to the sender role to find its next hop.
\begin{theorem}
    The broadcast-wait protocol, upon termination, guarantees that every IC node in the system reaches the sink with the least number of hops.
\end{theorem}

\begin{proof}
    We prove it by contradiction. Assume the protocol has terminated and the path from an IC node to the sink is not the shortest. Without loss of generality, we consider the scenario where hops \textsf{A}-\textsf{B} and \textsf{B}-\textsf{C} are on the path and \textsf{A} is also within the coverage of \textsf{C}. When \textsf{C} broadcasts its hop count, there are two cases: (1) Both \textsf{B} and \textsf{A} have received this message. In this case, \textsf{A} would use \textsf{C} as its next hop instead of \textsf{B} due to a shorter distance. Hence, both \textsf{A}-\textsf{B} and \textsf{B}-\textsf{C} would not be on the path. (2) Only \textsf{B} has received the message and \textsf{A} has not. In this case, \textsf{C} would follow the protocol and continue its broadcast until all its neighbors have received the message. Hence, the protocol has not been terminated. In either case, there is a contradiction. 
\end{proof}

\section{Low-Latency Message Forwarding}
\label{sec:routing}


In this section, we introduce a message-forwarding protocol to achieve low latency for packet delivery. First, we propose an efficient synchronization mechanism that allows each IC node to synchronize its working period (slot) with its next hop for communication in bounded time slots. Based on this synchronization mechanism, we then explain the message forwarding protocol, where each IC node caches the working time offset to its next hop to reduce repeated expensive synchronization overhead. 

\subsection{Working Time Synchronization}
\begin{figure}[!t]
    \centering
    \includegraphics[width=0.42\textwidth,page=4]{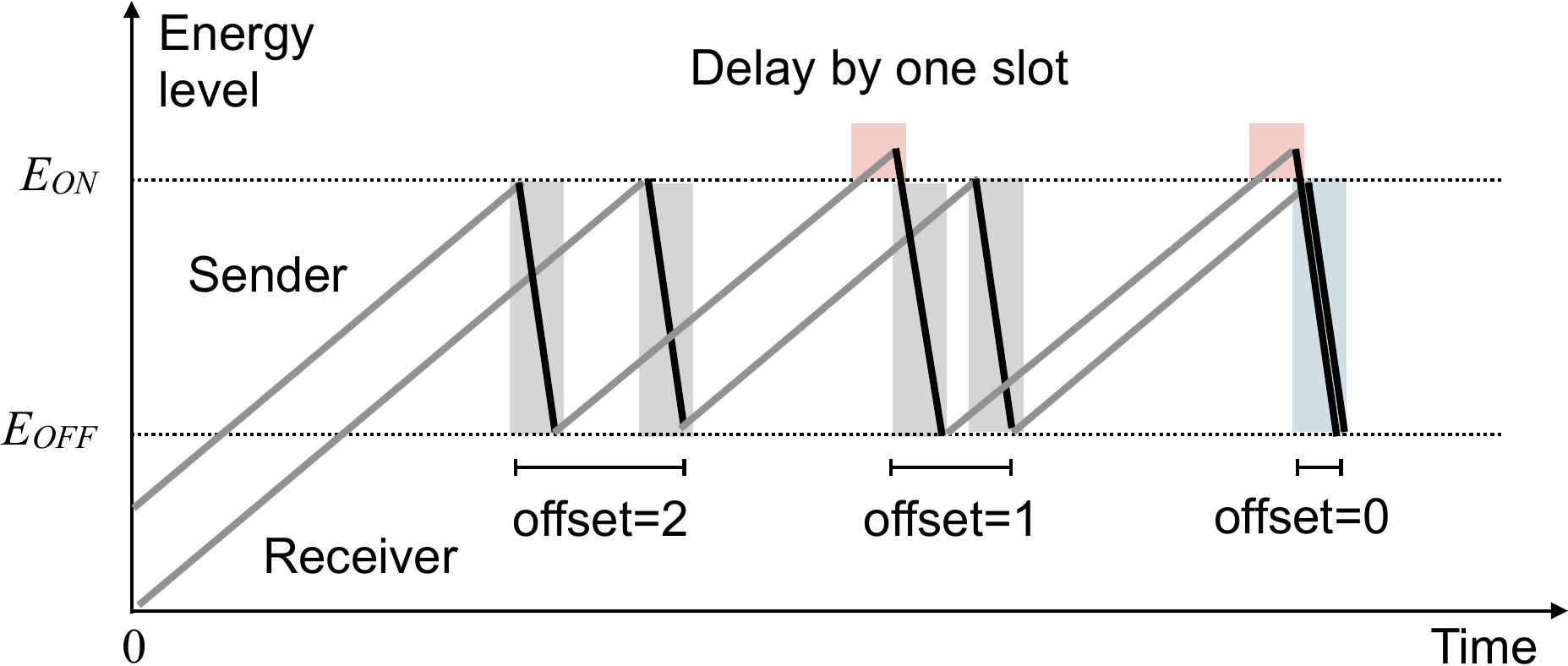}
    \caption{The working time synchronization process of two IC nodes as the sender and receiver respectively.}
    \label{fig:sync}
\end{figure}

For an IC node to send messages to its next hop (assuming the next hop is not the sink), the two IC nodes have to be in the working state in the same time slot. The main challenge consists in synchronizing the working time of the two IC nodes. 
We propose a simple yet effective synchronization mechanism where in every charging cycle, the sender delays its working time by one slot, as depicted in Figure~\ref{fig:sync}. 
This process continues until the working period of the sender is aligned with that of the receiver. 
The receiver is reactive and simply listens to the synchronization signals from the sender following its original charging cycles. 
This way, the sender walks through all the possible working time offsets between the two IC devices. 
Upon the alignment of working time, the sender will be able to send the message to the receiver successfully in all subsequent charging cycles. 

Since the working time offset between any two IC devices is upper-bounded by $t$, the maximum possible synchronization time is given by $t \cdot (t+1)$ where $(t+1)$ is the length of a charging cycle. 
Compared with existing randomized approaches like Find~\cite{2021-nsdi-find}, our synchronization mechanism ensures that each IC node can find its next hop in a provable bounded time while achieving comparable average synchronization time as with the randomized approaches, as later shown by our evaluation results in Section~\ref{sec:evaluation}.

\subsection{Message Forwarding with Pendulum-Sync}
\begin{figure}[!t]
    \centering
    \includegraphics[width=0.42\textwidth,page=5]{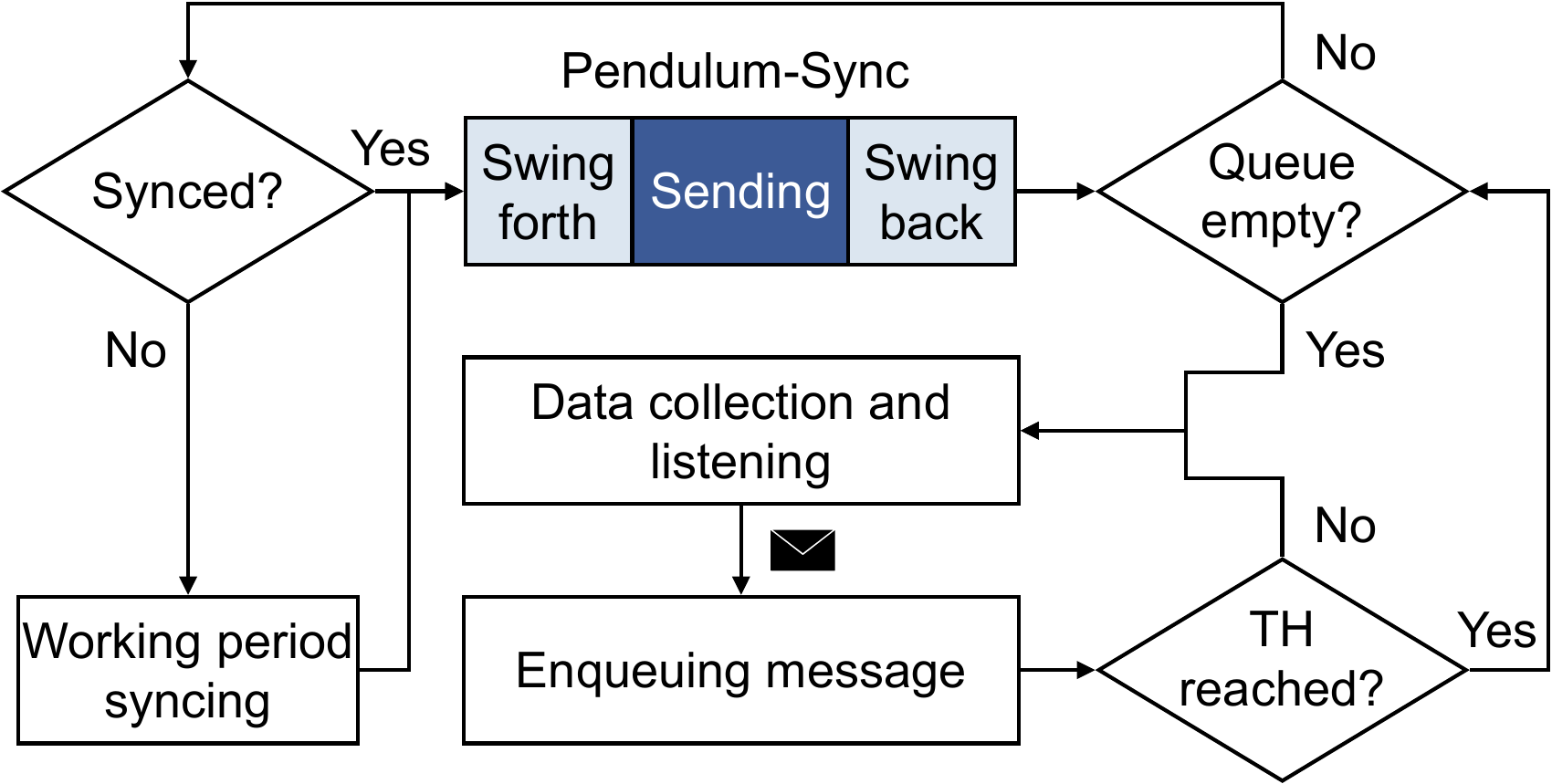}
    \caption{The workflow of an IC node during message forwarding, following the pendulum-sync-based protocol.}
    \label{fig:forwarding}
\end{figure}

Following the paths generated during topology construction, each IC node will receive messages from its previous hop(s) if any, and send and/or forward messages to its next hop on its path toward the sink. 
In this procedure, each IC node serves two roles alternately: sender and receiver. 
An IC node, when serving the sender role, first synchronizes its working time with its next hop (using the above synchronization mechanism). 
After that, the IC node can send messages to its next hop. 
An IC node with the receiver role keeps listening and receiving messages from its synchronized sender and puts the received messages into its sending queue. 
The IC node switches to the sender role to send out the messages cached in its sending queue to its own next hop when enough messages have been queued up. 
While the path leading to the sink for each IC node is known, each IC node still needs to synchronize with its next hop with respect to the working time before the message can be forwarded to its next hop. 
If this has to be done for every hop on the path, the time overhead will be prohibitive when the energy condition is poor.

To address this issue, we propose an efficient message-forwarding protocol based on a novel technique called \emph{pendulum-sync}. 
Our idea is inspired by a simple observation that the working time offset between IC nodes stays stable during a relatively long period of time under stable environmental conditions~\cite{2021-nsdi-find}. 
Hence, an IC node, once having figured out the working time offset to its next hop, can cache the offset locally in its non-volatile memory and reuse it for all subsequent message-forwarding operations, thus avoiding repeated synchronization. 
Our proposed message-forwarding mechanism exploits this observation and instructs each IC node to swing its working time forth and back using the cached offset to enable efficient message forwarding. 

Each IC node follows the workflow in Figure~\ref{fig:forwarding}. The IC node initially stays in the data collection and listening state to produce and/or receive messages. This procedure repeats until the sending queue length of the IC node reaches a pre-defined threshold. When the queue length reaches the threshold (denoted as TH in the figure), the node becomes a sender and starts sending messages in the sending queue to its next hop. To this end, the IC node first runs the working time synchronization mechanism to obtain the working time offset to its next hop and cache the offset value in its non-volatile memory. To send a message, the IC node runs the pendulum-sync protocol: swinging forth its working time to align with its next hop, sending the message, and swinging back to its original working time. The IC node repeats this produce until the sending queue becomes empty. After that, the IC node becomes a  receiver and starts data collection and listening again. 
In the following, we explain the protocol by detailing the sending and receiving procedures.
Although the collision rate is low due to the low duty cycle of each IC node, the cost is high when a collision happens. To further reduce collisions, each receiver sends an acknowledgment with a random delay, each sender sends its messages also with a random delay.

\subsection{Sending Procedure}
\begin{algorithm}[!t]
\small
\caption{IC node sending procedure}\label{algo:sending}
\begin{algorithmic}[1]
\State {$attempt\_{send} \gets 0$, $flag\_{match} \gets false$, $id\_next \gets -1$}
\State {$offset\_{forth} \gets 0$, $offset\_{back} \gets 0$}

\If{$queue.size == 0$}
    \State{Return to listening state by setting $offset\_{back}$}
\EndIf

\If{$attempt\_{send} == t + 1$}
    \State {$attempt\_{send} \gets 0$}
    \State{Return to listening state by setting $offset\_{back}$}
\EndIf

\While {$queue.size$ $!= 0$}
    \If{$flag\_{match} == false$}
        \State{$attempt\_{send} \gets attempt\_{send} + 1$}
        \State {$offset\_{forth} \gets offset\_{forth} + 1$}
        \State {$offset\_{back} \gets t + 1 - offset\_{forth}$}
        \State {$max\_messages = queue.size$}
    \EndIf
    \State {$message$ = $queue$.\texttt{pop}(), $message.dst \gets id\_next$}
    \If{$queue.size == max\_messages - 1$}
        \State{ $message.is\_start \gets true$}
    \EndIf
    \If{$queue.size == 0$}
        \State{$message.is\_end \gets true$}
    \EndIf
    \State {$ack =$ \texttt{broadcast}($message$)}
    \If{$ack$ and $ack.dst == self.id$}
        \State {$flag\_{match} \gets true$}
        \State {$id\_next \gets ack.src$}
    \Else
        \State {$queue$.\texttt{push}($message$)}
\EndIf
\EndWhile
\end{algorithmic}
\end{algorithm}


All IC nodes, except the edge IC nodes that will not serve as relay nodes, stay in the listening mode initially. 
An IC node, once having produced enough data, will start the sending procedure as listed in Algorithm~\ref{algo:sending}. 
Each IC node sets a limit for the number of sending attempts. 
Since in every attempt, the IC node delays its working time by one slot, the IC node needs at most $t+1$ attempts before it can find its next hop for communication. 
If the sending queue becomes empty or the sending operation has not succeeded before reaching the limit, the IC node swings back its working time offset and returns to the listening state (lines 3-7). 
For the case where the sending queue is not empty, if the current IC node has not been synchronized with its next hop, it will start the working period synchronization procedure by delaying its working time by one slot in every sending attempt until a success (lines 10-11). 
Meanwhile, the node records its current offset to its original working time so it can swing back later (line 12). 
For every sending attempt, the node dequeues a message from the queue and broadcasts the message, hoping to receive an acknowledgment from the next hop (lines 13-18). 
If an acknowledgment has indeed been received, the current node will be matched to the node which has sent the acknowledgment until all the messages in the queue have been sent out (lines 19-21); otherwise, the message will be enqueued to retry in the next attempt (line 23). 
Upon link/node failures, each sender searches for its next hop again by waiting for $(t+1) \times q_{max} + \Delta$ time slots, where $q_{max}$ is the maximum queue length and $\Delta$ is the tolerance waiting time. 


\subsection{Receiving Procedure}

\begin{algorithm}[!t]
\small
\caption{IC node receiving procedure}\label{algo:receiving}
\begin{algorithmic}[1]
\State {$id\_match \gets -1$, $time\_{wait} \gets 0$}

\While {$time\_{wait}$ $<= t+1$ or $queue.size == 0$}
    \State{$pacekt$ = \texttt{receive}()}
    \If{ $message$ is not $null$ }
        \If{ $message.src.hop > self.hop $ and \newline
                 \hspace*{5em} ($message.src.id\_next == self.id$ or \newline
                 \hspace*{5em} ($ message.src.id\_next == -1$ and  \newline
                 \hspace*{5em} $ id\_match == -1$))}
            \State {$queue.$\texttt{push}($message$)}
            \State {Create an acknowledge $ack$}
            \State {$ack.dst \gets id\_match$} 
            \If {$message.is\_start == true$}
                \State {$id\_match \gets message.src$}
                \State {$ack.dst \gets id\_match$}
                \State {\texttt{broadcast}($ack$)}
                \State {$time\_wait \gets 0$ }
                \State Continue
            \ElsIf {$message.is\_end == true$ }
                \State {$id\_match \gets -1$}
                \State {\texttt{broadcast}($ack$)}
                \State {Return to sending state}
            \Else
                \State {\texttt{broadcast}($ack$)}
                \State Continue
            \EndIf
        \Else
            \State{\texttt{drop}($message$)}
        \EndIf
    \EndIf
    
    \If { $message.src.id\_next != self.id$ }
        \State {$id\_match \gets -1$}
        \State {$queue.$\texttt{pop}()}
    \EndIf
    \State $time\_{wait} \gets time\_{wait} + 1$
\EndWhile
\If{$time\_{wait} > t+1$}
    \State {$time\_{wait} \gets 0$}
    \State{Return to sending state}
\EndIf
\end{algorithmic}
\end{algorithm}

An IC node keeps listening to receive data if its sending queue is empty or it has sent in a row of $t+1$ rounds without any response, following the procedure listed in Algorithm~\ref{algo:receiving}.
In each round, the IC node expects to receive a message (line 3). 
If the received message is legitimate, the message will be pushed to the sending queue (lines 5-6).
An acknowledgment will be created immediately which will be sent back to the sender of the just received message. 
If this is the first time receiving a message from this sender, the current node will lock with the sender so that no synchronization is needed for subsequent messages and no other nodes can send messages to the current node before this sender finishes (lines 9-12). 
At this point, the current node also restarts its counter for the receiving round so it can receive up to $t+1$ messages before turning into the sending state (line 13). 
If the received message is the last one from this sender, the current node will release the lock with the sender, since it may receive messages from other senders later (lines 15-17). 
The node also returns to the sending state to empty its sending queue (line 18). 
Illegitimate messages will be dropped by the node directly. 
It might happen that two nodes have received a message from the same sender, but the sender will just accept the acknowledgment from one of the nodes (depending on which acknowledgment arrives first). 
In the second round, the other node will notice that the sender has picked a different node as its receiver and thus will break the failed lock and drop the message that has been previously received (lines 24-26). 
If the node has been in the receiving state for too long, the node enters the sending state (lines 28-30).

\section{Performance Evaluation}
\label{sec:evaluation}

We implement \namex in the OMNeT++ simulator, which is a widely used discrete time simulator for network simulations and is convenient for validating the properties and performance of our proposed protocols. 
Our simulation communication protocol is based on the BLE Mesh protocol~\cite{2018-specification-BLEMesh}. BLE Mesh has high performance to avoid collisions with random delay and multiple channels. 
Yet, it is hard to build a stable connection directly between paired nodes due to the tiny working times of IC nodes.
Based on our implementation, we perform extensive simulation studies to evaluate the performance of the proposed protocols and mechanisms in \namex. 
In the following, we first explain our evaluation setups, and baselines, and then discuss the evaluation results in detail.

\subsection{Experimental Setups}
\label{sec:evaluation:setup}




We create two areas where IC nodes are scattered: \textit{square} (45m$\times$45m with the sink node located at the middle of an edge) and \textit{rectangle} (40m$\times$80m with the sink node located in the middle of a shorter edge). 
These two scenarios cover cases with different average hop counts in the network.
The sink node is not intermittently-powered and thus it can work at any time. 
To test the performance of RICS under different hop counts, we consider two cases, with 50 and 100 IC nodes, respectively. 
The IC nodes are randomly distributed in the area to simulate real-world conditions~\cite{2018-trustcom-esrq}, and work under the same ambient environment with charging cycles of the same length. 
The communication range of each IC node is set to 10m. 
We consider three energy conditions for all the IC nodes: \textit{good} ($t=5$ slots), \textit{median} ($t=120$ slots), and \textit{poor} ($t=500$ slots). 
These energy conditions are broad enough to cover a large variety of real-world scenarios~\cite{2019-sensys-shepherd}.  
The initial working time of each IC node is randomly drawn from $[0,t]$. 
Each IC node produces and sends out messages in a periodical fashion: one message is created in every charging cycle. 
The message is expected to be sent out in the next charging cycle.
We consider different network load cases where the IC node sends different rounds of messages as detailed in each experiment. 

\subsection{Evaluation Metrics and Baselines}
\label{sec:evaluation:baselines}


For evaluating the topology construction mechanism, we focus on the total time (in slots) required for the topology to be established. 
We also assess the practicality of the topology construction mechanism by setting the slot length to 1ms as a realistic number~\cite{2021-nsdi-find}. 
For evaluating the message forwarding mechanism, we focus on the per-message delivery time, e.g., the time spent between when the message is sent at its source node and when the message arrives at the sink node. 
In particular, we will look at the distribution of per-message delivery time. 
The performance of the message-forwarding mechanism is evaluated against the following baselines:
\begin{description}
\item[FXCS] Fixed continuous synchronization adopts a fixed next hop (generated by topology construction) for message forwarding as our proposal but does not adopt the caching of working period offsets on IC nodes, i.e., no pendulum-sync-based forwarding. 
\item[RNCS] Random continuous synchronization requires nodes to synchronize with their next hops repeatedly for message forwarding. Each node selects its next hop randomly and does not cache the working period offset for any synchronized next hop, so working period synchronization is required before every transmission. 
\item[OPPS] Opportunistic pendulum-sync approach allows each node to cache the working period offset for its next hop to avoid repeated working period synchronization. When a node sending messages cannot establish a match with its already discovered next hop, it will try to match with another node as its new next hop opportunistically. 
\end{description}
The above baselines are variants of our \namex with certain components disabled or slightly modified. Since \namex is the first of its kind, we regard our baselines as state-of-the-art.

\subsection{Performance of Topology Construction}
\label{sec:evaluation:perf-topo}

\begin{figure}[!t]
    \centering
    \includegraphics[width=0.48\textwidth]{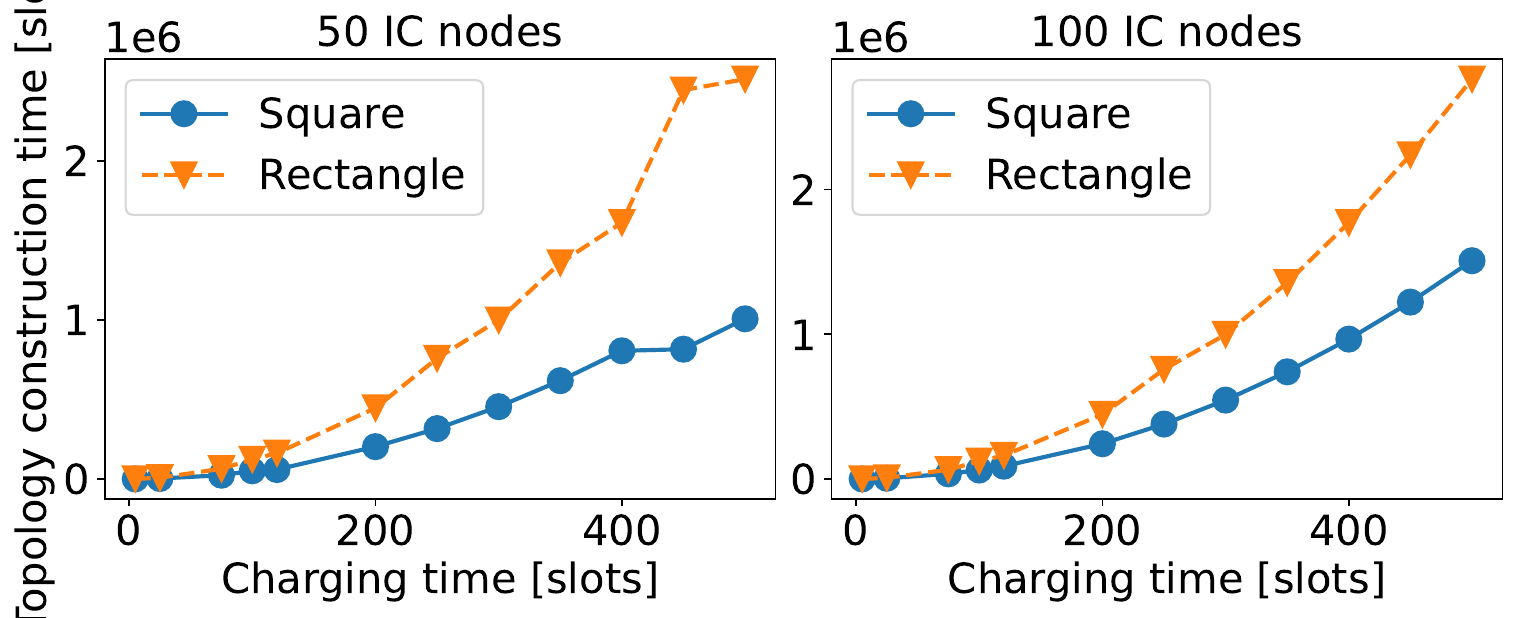}
    \caption{Topology construction time with \namex under different scenarios and varying numbers of IC nodes.}
    \label{fig:perf-topo}
\end{figure}

We first verify the correctness and performance of the topology construction protocol. 
By analyzing the logs of the experiments, we confirm that our proposed protocol establishes the topology correctly, i.e., all the IC nodes are able to obtain the shortest path to the sink node. 
To evaluate the performance, we measure the topology construction time under the two area scenarios and with 50 and 100 nodes respectively. 
We test the performance under varying energy conditions with charging times ranging from 5 to 500 slots. 
To avoid randomness, we repeat this experiment ten times and plot the average.

Fig.~\ref{fig:perf-topo} shows the results for the topology construction time. 
As we can see that with the increase in the charging time, the topology construction time increases accordingly. 
This is because, with a larger charging time, the working period synchronization and communication between nodes take longer to finish. 
We observe that it takes much shorter to construct the topology with the square scenario compared with the rectangle scenario as expected. 
This is due to that in the former the sink node is in the middle of the area and nodes generally have fewer hops to reach the sink. 
The increased hop count delays the information propagation in the broadcast process, leading to a longer finishing time.
With more IC nodes distributed in the same area, the topology construction takes slightly more time, but the difference is marginal since the average hop count is largely influenced by the area size and the communication range of the IC node and not much by the number of nodes in the area (considering that only the shortest path will be preserved by the protocol).

We also assess the practicality of our topology construction mechanism by substituting a time slot with 1ms, which is already quite conservative for real-world scenarios~\cite{2021-nsdi-find}. 
Under this situation, the topology construction takes less than one second in both area scenarios under good energy conditions (i.e., $t=5$ slots). 
In the worst case under very poor energy conditions (i.e., $t=500$), 
the topology construction can finish within 17min and 42min in the square and rectangle scenarios with 50 IC nodes, respectively. 
With 100 IC nodes, the topology construction can finish within 26min and 47min in the square and rectangle scenarios, respectively.
Note that the topology construction is only needed when the environment changes, which typically happens at the hour scale. 
Therefore, we conclude that our proposed topology construction mechanism is practical.

\subsection{Performance of Working Time Synchronization}
\label{sec:evaluation:perf-sync}

\begin{figure}[!t]
        \centering
        \includegraphics[width=0.48\textwidth]{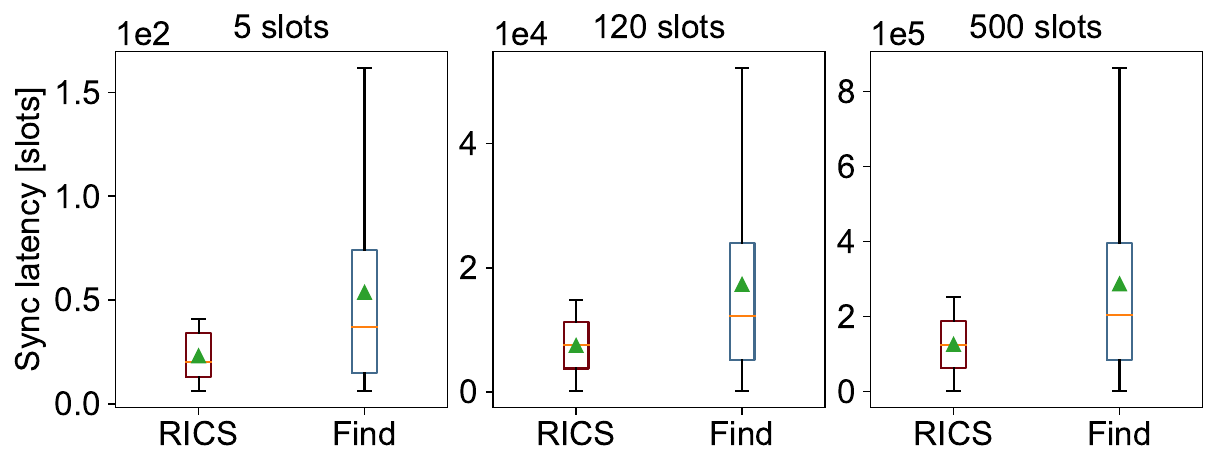}
        \caption{Synchronization latency (in slots) comparison between our proposed mechanism and Find~\cite{2021-nsdi-find}. The charging time $t$ is set to $\{5,120,500\}$ slots, representing good, median, and poor energy conditions, respectively.}
        \label{fig:cysync-find}
\end{figure}

We now evaluate the performance of the working time synchronization mechanism. We observe that our proposed mechanism while being simple is quite effective under different energy conditions (i.e., different charging times). 
To demonstrate that, we have conducted a simulation study in comparison with the randomized synchronization mechanism Find~\cite{2021-nsdi-find} which is considered the state of the art. 
We vary the charging time with $t \in \{5, 120, 500\}$ slots, which represent good, median, and poor energy conditions, respectively.
The working time of the two IC devices (relative to the system initialization time) is drawn uniformly at random from $[0, t]$. 
We repeat the experiment 10K times and compare the time it takes for the sender to discover the receiver, i.e., the synchronization latency. 
Fig.~\ref{fig:cysync-find} depicts the comparison results. 
As we can see that our proposed mechanism achieved comparable (and generally lower) synchronization latency on average when compared with Find, under all three considered energy conditions. 
More importantly, the latency variance of our proposed mechanism is much lower than that of Find. 
Note that Find is sensitive to the parameter setups specific to the underlying distribution of the working time offsets of IC nodes. 
With careful parameter tuning, Find may achieve better performance, although the improvement would be quite marginal as discussed in~\cite{2021-nsdi-find}.
In real-world scenarios, the synchronization latency is quite acceptable. 
Assuming a time slot of $1$ms, our proposed mechanism can complete the discovery of two IC devices in $0.022$s, $7.457$s, and $125.028$s, under the three energy conditions, respectively.

\begin{figure*}
    \centering
    \begin{subfigure}[b]{0.24\textwidth}
        \centering
        \includegraphics[width=\textwidth]{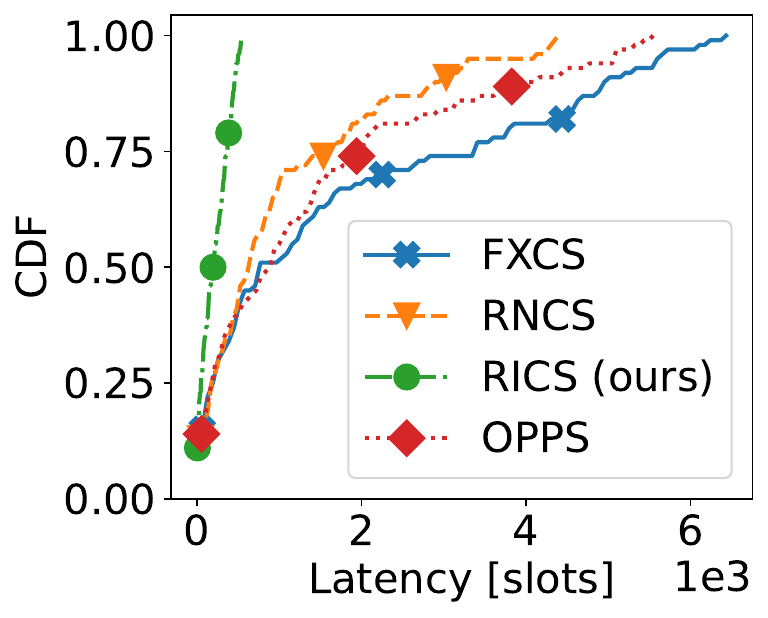}
        \caption{}
        \label{fig:perf-overall}
    \end{subfigure}
    \hfill
    \begin{subfigure}[b]{0.24\textwidth}
        \centering
        \includegraphics[width=\textwidth]{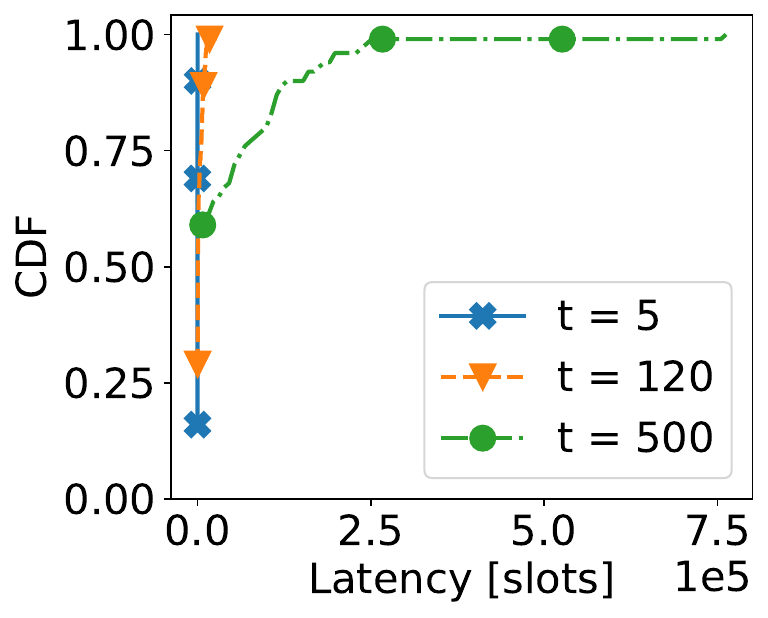}
        \caption{}
        \label{fig:perf-charging}
    \end{subfigure}
    \hfill
    \begin{subfigure}[b]{0.24\textwidth}
        \centering
        \includegraphics[width=\textwidth]{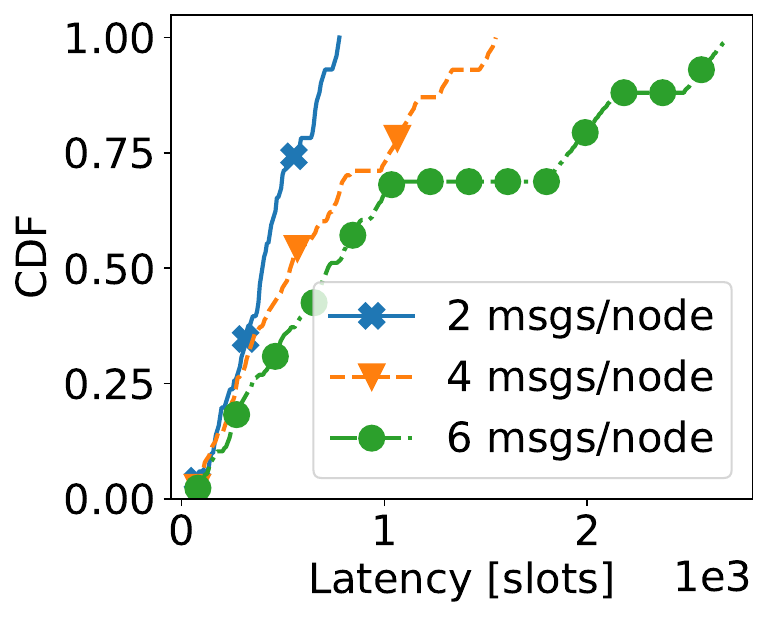}
        \caption{}
        \label{fig:perf-load}
    \end{subfigure}
    \hfill
    \begin{subfigure}[b]{0.24\textwidth}
        \centering
        \includegraphics[width=\textwidth]{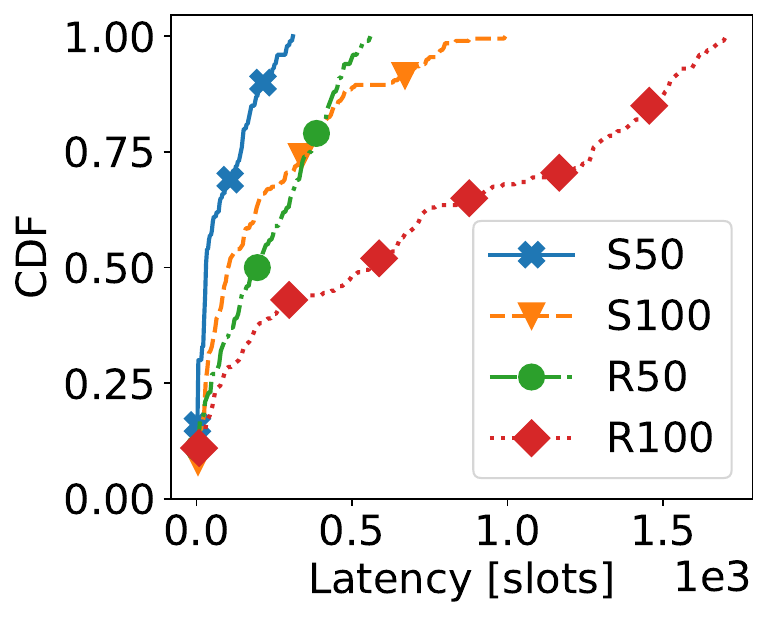}
        \caption{}
        \label{fig:perf-size}
    \end{subfigure}
    \caption{End-to-end per-message delivery time CDF of \namex: (a) compared with the baselines, (b) under varying charging time $t=50,120,500$, (c) under varying network loads, and (d) under different network sizes and shapes. 
    }
\end{figure*}

\subsection{Overall Performance of \namex}
\label{sec:evaluation:overall}



We now evaluate the overall performance of \namex by comparing it with the baselines. 
We choose the square scenario with 50 IC nodes. 
The charging time of the IC node is set to $t=50$ slots.
Each IC node is instructed to send two rounds of messages. 
We run \namex and the baselines separately and compare the message delivery time.

Figure~\ref{fig:perf-overall} depicts the cumulative distribution function (CDF) for the per-message end-to-end latency. As we can see that without the pendulum-sync-based forwarding protocol, the random approach (RNCS) has generally better performance than the fixed approach (FXCS) as expected. This is because without pendulum-sync, the synchronization overhead is large and the approach with a random next hop can better leverage the path redundancy in the network to reduce the delivery time, similar to the cooperative relaying idea explored in~\cite{2018-tecs-predic}. 
With pendulum-sync, \namex performs significantly better than all the baselines including OPPS which also adopts pendulum-sync. 
This is because OPPS employs opportunistic probing, trying to reap the benefit of path redundancy, but such probing adds drastic synchronization overhead, which over-weights the benefits provided by opportunistic probing. 
Overall, we can see that \namex achieves significantly lower (by orders of magnitude) message delivery time than all the baselines, proving the efficiency of the pendulum-sync-based message forwarding as well as \namex as a whole. 

\subsection{Impact of the Energy Condition}
\label{sec:evaluation:impact-energy}



We now study the impact of the energy condition (dictating the charging time of IC nodes) on the performance of \namex. Similar to the previous study, we consider the square scenario with 50 IC nodes. Each IC node sends two messages in every charging cycle. We vary the charging time of the IC node with $t=\{50, 120, 500\}$ and record the per-message delivery time. 

Figure~\ref{fig:perf-charging} shows the CDF for the per-message end-to-end latency. We can observe that with the increase in the charging slot length, the per-message delivery time increases accordingly. This is reasonable since the working period synchronization takes longer with longer charging slots. Moreover, with longer charging slots, per-hop latency will also increase since each IC node will have to wait for a long to be charged before it can be activated for communication. 
Overall, \namex achieves excellent performance under relatively good energy conditions, i.e., $t=\{5, 120\}$, but suffers a long tail message delivery time under poor energy conditions, i.e., $t=500$. This is because, under poor energy conditions, it takes longer for IC nodes to synchronize on every hop in the worst case, adding to the end-to-end delivery time.

\subsection{Impact of the Network Load}
\label{sec:evaluation:impact-load}



We now evaluate the impact of the network load on the performance of \namex. Again, we choose the square scenario with 50 IC nodes. The charging time of the IC node is set to $t=50$ slots. We vary the number of rounds each IC node sends messages and record the per-message delivery time. 

Figure~\ref{fig:perf-load} shows the CDF of the per-message delivery time under different network loads. As shown in the figure, when the network load increases (i.e., more messages to be sent over the network in a given amount of time), the per-message delivery time increases, as expected. The latency increase can be attributed to the longer queueing time of messages at IC nodes due to the limited bandwidth resources on the shortest paths produced by the topology construction. We conjecture that opportunistically leveraging the path redundancy in the network may improve the network throughput and thus alleviates this issue. However, as shown in Figure~\ref{fig:perf-overall}, the extra synchronization overhead needs to be controlled carefully in order to achieve the gain. 

\subsection{Impact of the Network Size and Shape}
\label{sec:evaluation:impact-size}


Finally, we study the impact of the network size and shape on the performance of \namex. We set the charging time of all IC nodes to $t=50$ and instruct each IC node to send two rounds of messages. We consider the following four scenarios: a square area with 50 nodes (S50), a square area with 100 nodes (S100), a rectangle area with 50 nodes (R50), and a rectangle area with 100 nodes (R100).

Figure~\ref{fig:perf-size} illustrates the CDF of the per-message delivery time under the above four scenarios. Comparing the results under the same shape, e.g., S50 vs. S100 and R50 vs. R100, we can see that with the increase of the network size, the per-message delivery time increases. This is mainly because, with a higher number of nodes, more messages need to be delivered on the network considering that each IC node generates messages. Under the same network size, we can see that the scenarios with a square area generally have lower per-message delivery times. This can be explained by the fact that with the square scenario, the average hop count for each IC node to reach the sink is larger. This is attributed to that the sink is located in the middle of the square area and at the edge of the rectangle area.

\section{Limitations and Discussion}
\label{sec:discussion}


\textbf{Heterogeneous charging cycles.} Existing works often assume that the charging cycles of IC nodes are the same~\cite{2021-nsdi-find}, as we also do in our work. 
This assumption may not always be true in the real world, making the working time synchronization extremely challenging.
Typically, the assumption holds for the charging cycles of IC nodes dispersed in a small area but would become inapplicable in large areas. 
One approach to tackle this problem is to partition the large area into smaller blocks and assume a common charging cycle for each of these blocks. 
However, coordinating these blocks in a completely distributed manner is still a challenging issue. 

\textbf{Redundant paths.} \namex only leverages a single next hop for packet routing on the network. 
While being efficient under low traffic, it may lead to hot spots in the network when the traffic increases. 
We have verified that a naive multi-path approach that leverages multiple next hops in an opportunistic manner does not help due to the introduced synchronization overhead. 
Therefore, a more sophisticated traffic-spreading mechanism is required to spread the traffic over the nodes in the system and achieve load balance in general. 
The problem is hard to solve in a totally distributed environment. 

\section{Conclusions and Future Work}
\label{sec:conclusions}
Intermittently-powered sensing systems relying completely on battery-free devices have received a lot of attention recently, due to their huge potential in enabling perpetual sensing. 
However, a key problem of data delivery across IC devices has not been addressed, hindering the deployment and application of intermittently-powered sensing systems.
In this paper, we propose \namex, an efficient routing scheme tailored for delivering data across IC devices efficiently. 
\namex addresses this routing problem with two key designs: a fast topology construction protocol that allows each IC node to obtain a least-hop path towards the sink node based on coordinated flooding and an efficient message forwarding protocol based on pendulum-sync to avoid expensive repeated working time synchronization. 
Our evaluation based on an OMNeT++ implementation and experiments under varying scenarios and setups shows that \namex achieves its goal and outperforms the baselines significantly. 
In future work, we will generalize \namex to non-uniform and even varying charging times. We also expect to test our protocol in a real-world testbed built with multiple IC devices we are currently building. 

\section*{Acknowledgment}
This work has been partially funded by the Dutch Research Council (NWO) grant OCENW.XS22.1.135.
Gaosheng Liu is funded by the China Scholarship Council (CSC) fellowship.

\bibliographystyle{IEEEtran}
\bibliography{refs}

\end{document}